\documentclass[useAMS,usenatbib]{mn2e}
\usepackage{amsmath,graphicx}

\newcommand{\kms}{$\rm\thinspace km\ s^{-1}\,$}

\title[Reflection-dominated X-ray spectra of NLS1: Mrk 478 and EXO 1346.2+2645]{Reflection dominated X-ray spectra of Narrow Line Seyfert 1 Galaxies: Mrk 478 and EXO 1346.2+2645}
\author[A. Zoghbi et. al]{A. Zoghbi$^{1}$\thanks{E-mail:
azoghbi@ast.cam.ac.uk}, A. C. Fabian$^{1}$ and L. C. Gallo$^{2}$\\
$^{1}$Institute of Astronomy, Madingley Road, Cambridge CB3 0HA\\
$^{2}$Department of Astronomy and Physics, Saint Mary's University, Halifax, NS B3H 3C3, Canada}
\begin{document}

\pagerange{\pageref{firstpage}--\pageref{lastpage}} 

\maketitle

\label{firstpage}

\begin{abstract}
Multi-epoch \emph{XMM-Newton} spectra of the two Narrow Line Seyfert 1 (NLS1), Mrk 478 and EXO 1346.2+2645, are presented. The data were fitted with different models, including the relativistically-blurred reflection model {\sc reflionx}, which was found to give a good description of both spectra in all epochs. The two sources are reflection dominated with the illuminating continuum hidden from view. This can be explained either in terms of a corrugated disc or strong gravitational light bending effects. The emission from these two sources comes from a very small region (few gravitational radii), very close to a rapidly spinning black hole.
Spectral variability analysis show that both sources have constant RMS spectra, and constant hardness ratios in all epochs, in agreement with a single spectral component. The constancy the spectrum between high and low states rules out alternative models such as partial covering.
\end{abstract}

\begin{keywords}
galaxies: active -- galaxies:Seyfert -- galaxies:nuclei -- galaxies: individual: Mrk 478, EXO 1346.2+2645 -- X-rays: galaxies.
\end{keywords}

\section{Introduction}

The first study of Narrow Line Seyfert 1 galaxies (hereafter NLS1) as a class was done by \cite{1985ApJ...297..166O}. The sources were identified as having: (1) Balmer lines only slightly broader than the forbidden lines such as [OIII], [NII] and [SII], with widths less than 2000 \kms, and (2) the line ratio [OIII]$\rm{\lambda5007/H\beta < 3}$ (\citealt{1989ApJ...342..224G}). The latter defines them as Seyfert 1 rather than 2. In addition, they show strong FeII emission lines and iron emission from higher ionisation states.

The importance of NLS1 became apparent after the discovery of the first Principle Component Analysis eigenvector (E1) of Seyfert 1s. This involves anti-correlations between the width of the Balmer lines and the strength of $\rm{FeII}$ emission line (\citealt{1992ApJS...80..109B}, see a review by \citealt{2000ARA&A..38..521S}). An additional anti-correlation was also found between ROSAT spectral index (0.1-2.4 keV) and the FWHM of the H$\beta$ line (\citealt{1996A&A...305...53B}). NLS1s occupy an extreme corner of this correlation parameter space, with small line widths, strong FeII emission and steep X-ray photon indices. These properties seem to be driven by small black hole masses and high Eddington ratios (\citealt{2000ARA&A..38..521S,2002ApJ...565...78B,2004AJ....127.1799G})).

\begin{table*}
 \centering
 \begin{minipage}{\textwidth}
  \caption{\emph{XMM-Newton} observation log.}
\label{tab:obs}
  \begin{tabular}{@{}llccccccc@{}}
  \hline
Source	&	z & Galactic absorption	&Obs.&	Obs.	&	Date	&	PN net$^b$&Average PN (0.2--10 keV)\\ 
Name	&	&	N$\rm{_H}$($\times 10 ^{20} \rm{cm}^{-2}$) $^a$ 
					&num.&	ID		&		&	exposure (ks)&count rate (counts/s)\\ \hline

Mrk 478 &	0.079&	1.03		&1&	0107660201&		23/12/2001&	10.1	&7.5\\	
	&	&			&2&	0005010101&		01/01/2001&	14.5	&5.2\\
	&	&			&3&	0005010201&		04/01/2001&	7.2	&5.3\\
	&	&			&4&	0005010301&		07/01/2001&	18.2	&2.8\\
EXO 1346.2+2645&0.0589&	1.19		&1&	0097820101&		26/06/2000&	41.9	&1.2\\
	&	&			&2&	0109070201&		13/01/2003&	45.7	&0.3\\
\hline
\multicolumn{8}{l}{$^a$ from \cite{1990ARA&A..28..215D}.}\\
\multicolumn{8}{l}{$^b$ after background flare subtraction.}
\end{tabular}
\end{minipage}
\end{table*}

At soft X-ray energies (less than $\sim1$ keV), NLS1 galaxies often show an excess above the extrapolation of a simple power-law from higher energies. This soft excess can usually be fitted by a blackbody with temperatures of $\sim$0.1--0.2 keV (\citealt{2004MNRAS.349L...7G}). The fact that the temperature is constant for different black hole masses, is suggestive of an atomic origin for the soft excess. \cite{2004MNRAS.349L...7G} suggested that relativistically smeared, partially ionised absorption can lead to an apparent soft excess below $\sim0.7$ keV. The drawback of this is that in order to produce the smooth spectrum, extreme outflow velocities ($\sim c$) are required (\citealt{2007MNRAS.381.1413S}). Alternatively, the soft excess can be explained in terms of relativistically blurred reflection (\citealt{2006MNRAS.365.1067C}), where the soft excess is the result of highly blurred emission lines resulting from the reflection off cold gas (\citealt{2005MNRAS.358..211R}).

In the hard band ($\sim$1--10 keV), NLS1 galaxies show in general a steeper photon index compared to `\emph{normal}' Broad Line Seyfert 1 , which is attributed to Compton cooling of the corona that emits the hard X-rays (\citealt{1997MNRAS.285L..25B}). A correlation was found between the spectral index and the Eddington ratio in samples that include NLS1 (\citealt{2004AJ....127.1799G,2008ApJ...682...81S}), and points towards the fact that NLS1 accrete at high rates. Another feature that is sometimes observed in NLS1 is a spectral drop around 7 keV, first observed in the source 1H 0707-495 (\citealt{2002MNRAS.329L...1B}), and later in other sources (e.g. IRAS 13224-3809, \citealt{2003MNRAS.343L..89B}). The feature appears to change in time (\citealt{2004MNRAS.353.1064G} for 1H 0707-495) and is not seen in all NLS1 AGN. Often it may appear as spectral curvature around 7 keV (e.g \citealt{2006MNRAS.368..479G}). The explanation of this drop is still not clear. Two models that have been suggested, both giving good fits to the data, are partial covering and reflection. 

In the partial covering model, the central source is covered by a patchy absorbing cloud. The absorption produces the sharp edge at high energies, while a fraction of the emission leaks through to produce the soft part of the spectrum. This, however, requires an excessively large iron abundance of $\sim30\times$ solar (\citealt{2002MNRAS.329L...1B}, see however \citealt{2004PASJ...56L...9T}). It should be noted that partial covering provides good fit for other sources where the spectral drop is not seen or not as sharp (e.g. \citealt{2007ApJ...668L.111G}). On the other hand, partially ionised reflection off an optically thick material (an accretion disc) can explain the spectra very well (\citealt{2004MNRAS.353.1071F}). In this model, the spectral drop around 7 keV is the blue wing of a relativistically broadened iron line.

In this paper we explore the spectra of two NLS1 using multi-epoch observations with \emph{XMM-Newton}. The paper is organised as follows. In section 2, a description of the observations and data reduction is presented. Spectral analysis and results are given in section 3, with subsections discussing individual sources. A study of the spectral variability is presented in section 4, and the implications are discussed in section 5.

\section{Observations \& Data Reduction}
The data presented here for the two sources Mrk 478 and EXO 1346.2+2645 consist of several observations which are detailed in Table \ref{tab:obs}. Some of the data have been published before as part of samples (e.g. \citealt{2004A&A...422...85P}, \citealt{2006MNRAS.365.1067C}, \citealt{2006MNRAS.368..479G}), but the majority is presented properly for the first time here.

The Observation Data Files (ODFs) were retrieved from the archive and processed in the standard way using \emph{XMM-Newton} Science Analysis System ({\sc sas v7.1.0}). The EPIC cameras were operated in large window mode in the first Mrk 478 observation (0107660201) and in small window mode in the last three.  EXO 1346.2+2645 was observed in full frame mode. The data were corrected for hot, dead and flickering pixels. Good Time Intervals (GTI) were produced in the standard way ({\sc sas} user guide \footnote{http://xmm.esa.int/external/xmm\_user\_support/document-ation/sas\_usg/USG/node57.html}). The most affected were the first and third observations of Mrk 478 (see Table \ref{tab:obs}). The pattern statistics were checked for possible pileup (using {\sc epatplot} in {\sc sas}), this was found to be present in the first Mrk478 observation. To correct for it, the central region of the source was excluded from the analysis. The pattern selection used is {\sc pattern} $\leq4$ for EPIC-PN and {\sc pattern} $\leq12$ for EPIC-MOS.\\
Sources spectra were extracted from circular regions 35 arcsec across centred on the source, and the background from regions on the same chip and away from the source. The spectra were then grouped such that each bin has a minimum of 20 counts. EPIC responce matrices were generated using the tasks {\sc rmfgen} and {\sc arfgen} in {\sc sas}.
It should be noted that EXO 1346.2+2645 is close in sky position to the cluster Abell 1795, and different background regions have been tested to make sure the spectrum of the source is not contaminated. Also, for the second observation of EXO 1346.2+2645 (0109070201), the calibration reported the central pixels of the source as bad so they were excluded from the analysis.
Spectral fitting was performed using {\sc xspec v12.4.0} (\citealt{1996ASPC..101...17A}). All quoted errors on the model parameters correspond to a 90 \% confidence level for one interesting parameter (i.e. a $\bigtriangleup\chi^2 = 2.71$) in the rest frame of the source.
\section{Spectral Analysis}
As a start, a simple power-law was fitted to each source in the energy range 0.3-10 keV. Galactic absorption was included using the column densities shown in Table \ref{tab:obs}. With multi-epoch observations, all fit parameters were tied except for a multiplicative constant. The PN and MOS data were fitted separately and they gave consistent results. In what follows (unless otherwise stated) only the PN data are used.
The simple power-law fit was not adequate, giving $\chi^2/d.o.f$ 3236/1760 and 903/787 for Mrk 478 and EXO 1346.2+2645 respectively. The soft band fitted well leaving most of the residuals at higher energies ($> 4$ keV).

\begin{figure*}
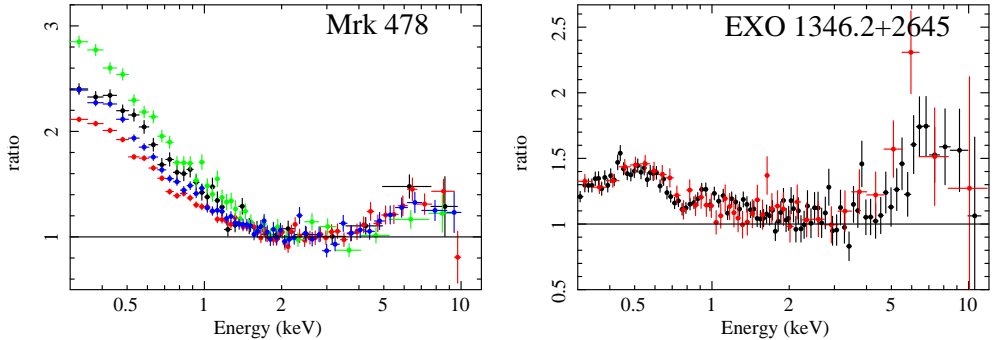

\centering
 \begin{tabular}{cc}
  \includegraphics[height=180pt,clip ,angle=270]{mrk_1_re.ps}&\includegraphics[height=180pt,clip,angle=270]{exo_1_re.ps}\\
 \end{tabular}
\caption{Data/model ratio of an absorbed power-law fit to the EPIC pn in the range 1.5-4.5, 8.5-10 keV extrapolated over the entire 0.3 to 10 keV for the two sources (the power-law was multiplied by a constant to account for flux variations). Different colours represent the different observations as shown in Table \ref{tab:obs}. Data have been re-binned for display.}
\label{fig:ratios}
\end{figure*}

The data were then fitted in the range 1.5-4.5 and 8.5-10 keV. This excludes the soft band where the soft excess dominates the spectrum and also excludes the region around the iron complex. A good fit was found with $\chi^2/d.o.f$ 674/685 and 222/289 for the two sources respectively. Fig. \ref{fig:ratios} shows the ratio of data/model when extrapolated over the 0.3-10 keV energy range. This clearly shows the soft excess emission below $\sim1.5$ keV. Also, it shows residuals around $\sim6-7$ keV that might be caused by the presence of iron emission.
\begin{figure}
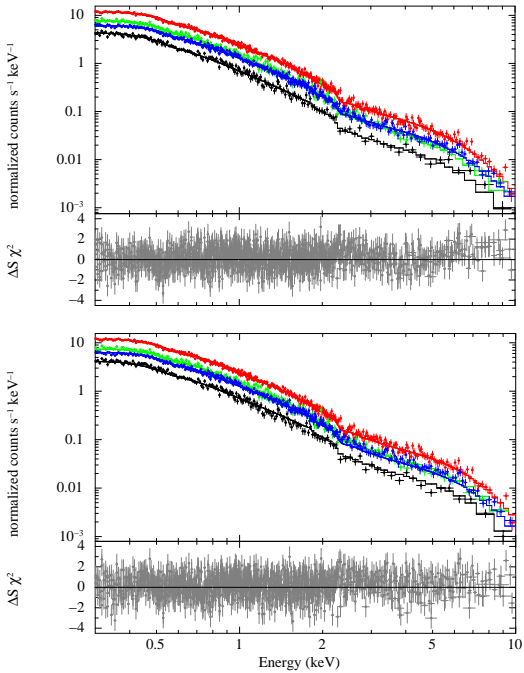

 \begin{tabular}{c}
  \includegraphics[height=200pt,clip ,angle=270, viewport=0 0 450 770]{mrk_ref_fit.ps}\\ \includegraphics[height=200pt,clip ,angle=270]{mrk_pc_fit.ps}
 \end{tabular}
\caption{The best fit models for Mrk 478 in the range 0.3-10 keV with residuals in terms of sigmas. \emph{top}: a single component reflection-dominated model. \emph{bottom}: a partial covering model with a blackbody and a partially covered power-law. \emph{Both} models include Galactic absorption, and the different colours represent the four observations available. Data have been re-binned for display.}
\label{fig:mrk_2_3}
\end{figure}

To find a phenomenological description of the data, each source will be looked at individually starting by adding a multi-colour disc blackbody ({\sc diskbb} in {\sc xspec}).

 \subsection{Mrk 478} Adding the blackbody component improved the quality of the fit significantly ($\bigtriangleup\chi^2 = $1041 for 2 d.o.f) compared to the simple power-law, the temperature of the blackbody was $123^{+2}_{-3}$ eV with a power-law index of $\Gamma = 2.41\pm0.02$ (compared to $89\pm0.12$ eV and $1.6\pm0.05$ respectively found with the \emph{ASCA} data fitted between 0.6 and 10 keV, \citealt{1999MNRAS.309..113V}). Allowing the temperature and $\Gamma$ to change between observations changes their values by 5 and 3 percent respectively, and improves the fits by $\bigtriangleup\chi^2 \sim$ 17 per additional free parameter. This might imply some minor variability in the source between observations. This fit, however, leaves some residuals at high energies, which can be fitted by a Gaussian line with rest frame energy of $E = 7.03^{+0.47}_{-0.40}$ keV and $\sigma = 1.59^{+0.42}_{-0.35}$ keV, and an equivalent width of EW = $1.36^{+0.29}_{-0.25}$ keV.\\
This broad line implies that relativistic effects are at play. Relativistically broadened lines {\sc diskline} (\citealt{1989MNRAS.238..729F}) and {\sc Laor} (\citealt{1991ApJ...376...90L}) emitted around a Schwarzschild and Kerr black hole, respectively, both, gave comparable fits (reduced $\chi^2$ of 1.07). The inner radii are $r_{\rm in}=6.0^{+1.1}_{-0.0}$, $r_{\rm in}=8.2^{+2.9}_{-1.9}$ gravitational radii ($r_{\rm g}=GM_{\rm BH}/c^2$), and line energies are $6.20^{+0.55}_{-0.54}$, $6.49^{+0.30}_{-0.13}$ keV, respectively, consistent with iron emission. These arise from the reflection of the incident power-law off the accretion disc, which implies the existence of a reflection continuum that has not been accounted for so far.

To model it, the high resolution version of the constant density reflection model {\sc reflionx} of \cite{2005MNRAS.358..211R} was used, blurred with {\sc kdblur} kernel of \cite{1991ApJ...376...90L} to account for relativistic effects in the vicinity of the black hole. In this model, an optically thick disc is illuminated with a power-law, producing florescence lines and continuum emission. Its parameters include the Fe abundance, ionisation parameter $\xi$ (defined as $\xi=4\pi F_{\rm tot}/n_H$ where $F_{\rm tot}$ is the total illuminating flux, and $n_H$ is the density of the reflector) and the incident power-law index $\Gamma$, while the {\sc kdblur} assumes a power-law emissivity and sharp inner and outer radii of the accretion disc\footnote{{phabs(kdblur(atable\{reflionx.mod\})+powerlaw)} in {\sc xspec}.} ($r_{\rm in}\ and\ r_{\rm out}$). The outer radius was frozen at 400 $r_{\rm g}$, as very little emission is expected outside this radius, while all other parameters were allowed to vary. The fit was very good with reduced $\chi^2$ of 1.017 for 1752 d.o.f.

In order to produce the smooth soft excess, the emission has to be highly blurred coming from a region very close to the black hole. The fits gave an inner radius of $r_{\rm in}=1.23^{+0.06}_{-0.00}\ r_{\rm g}$ (the inner most stable circular orbit of a Kerr black hole), with a very narrow emission region implied by the emissivity index peaking at the maximum value allowed ($10.0^{+0.0}_{-1.4}$).\\
The fact that some parameters ($r_{\rm in}$ and emissivity index) peak at the limits might be problematic, freezing the emissivity index at a lower value of 9 and 6 does not effect the other parameters, giving inner radii of $1.31^{+0.02}_{-0.06}~r_{\rm g}$ and $1.47^{+0.08}_{-0.02}~r_{\rm g}$ respectively. The top panel of Fig. \ref{fig:mrk_2_3} shows the data and residuals for the reflection model. The fit parameters are presented in Table \ref{tab:fits}.
\begin{table*}
\caption{The parameters of the best-fit reflection models.}
\label{tab:fits}
 \centering \scriptsize
\begin{tabular}{@{}lcllllllll@{}}
\hline
 Source	&Obs.&Em.	&$r_{in}$($r_g$)&Incl.($^{\rm{o}}$)	& Fe abun.& $\Gamma^{a}$	&$\xi^{a}$(erg cm s$^{-1}$)& Ref. Frac.$^{b}$&$\chi^2/\rm{d.o.f}$ \\
&Num&Ind.&&&($\times$solar)&&&&\\
\hline
Mrk 478	&1&6$^c$		&$1.47^{+0.08}_{-0.02}$&$34\pm2$&$0.36\pm0.03$&$2.43\pm0.02$&$1000^{+65}_{-208}$& $0.95\pm0.05$&1816/1754\\
&2&&&&&		$2.35\pm0.02$&	$971^{+70}_{-143}$& $0.81\pm0.09$\\
&3&&&&&		$2.47^{+0.02}_{-0.04}$&$1168^{+194}_{-104}$& $0.88\pm0.11$\\
&4&&&&&		$2.29\pm0.01$&$746^{+100}_{-76}$& $0.89\pm0.11$\\

EXO 1346&1&$3.9^{+1.1}_{-0.4}$&$1.6^{+0.6}_{-0.3}$&$36\pm8$&$0.34^{+0.05}_{-0.02}$&$2.17^{+0.02}_{-0.03}$ &$1202^{+120}_{-171}$& $0.83\pm0.09$&1255/1306 \\
&2&&&&&&&$0.96\pm0.04$\\
\hline
\multicolumn{8}{l}{$^a$ for Mrk 478, these parameters were allowed to vary between observations, and were tied for EXO 1346.2+2645}\\
\multicolumn{8}{l}{$^b$ calculated as the ratio of the flux in the reflection component to the total flux.}\\
\multicolumn{8}{l}{$^c$ parameter is frozen.}
\end{tabular}
\end{table*}

\begin{figure}
\centering
  \includegraphics[height=140pt,clip ]{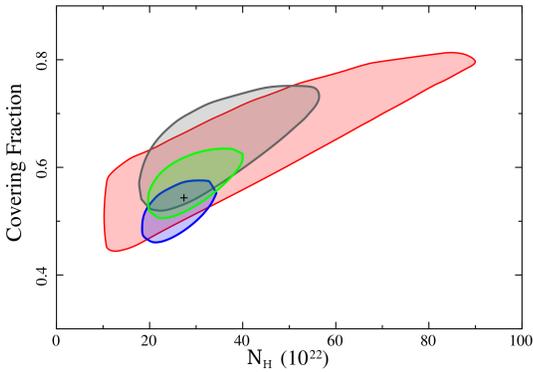}
\caption{A contour plot for the 90\% limits of the column density and covering fraction for the partial covering model for Mrk 478. The cross marks the best values for the combined data. The four colours represent the four epochs.}
\label{fig:mrk_pc}
\end{figure}
The normalisation of the power-law has a zero lower limit, which might indicate that the power-law is not required. A reflection-only model fits the data equally well (reduced $\chi^2/$d.o.f of 1.017/1753). The fit parameters remained almost the same, with the values of $\Gamma$ and $\xi$ changing by about 2 and 20 percent respectively when the power-law is removed (comparable to the errors in those values). Limits on the reflection fraction are shown in Table \ref{tab:fits}.

The other model discussed in the literature is partial covering. A good fit to the data was found using a continuum of a power-law and a blackbody with a single partial covering component ($\Gamma=2.66\pm0.02$, and $kT=112\pm20$ keV). The reduced $\chi^2$ was 1.006 for 1754 d.o.f. The data/residuals plot is shown in bottom panel of Fig. \ref{fig:mrk_2_3}. All parameters were tied between observations except for the power-law index and the normalisations. The model has an absorption column density of $2.7^{+2.9}_{-1.1}\times 10^{23}$ cm$^{-2}$, iron abundance of $0.2^{+0.7}_{-0.2}\times$ solar and a covering factor of $0.54\pm0.03$. Allowing the parameters to vary between observations (column density and covering fraction) only changes the size of the errors reflecting the quality of the individual observations. Fig. \ref{fig:mrk_pc} shows the 90 percent contour limits on the values of the column density and covering fraction. The coverer parameters remain constant between epochs.

\begin{figure}
\centering
  \includegraphics[height=220pt,clip ,angle=270]{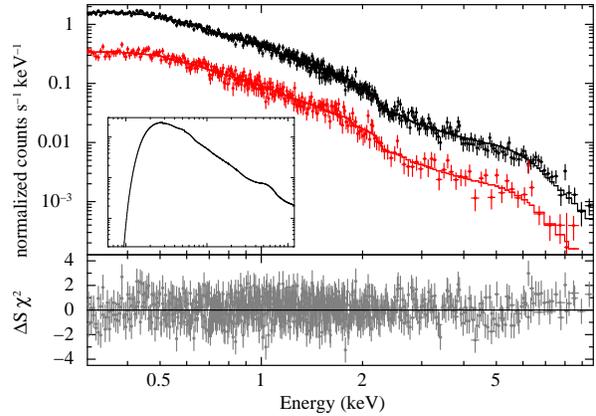}
\caption{The best fit model for EXO 1346.2+2645 in the range 0.3-10 keV with a single component reflection-dominated model, with residuals in units of sigma. Data have been re-binned for display, colours represent different observations (and detectors: PN and MOS) \emph{inset}: The model used to fit the data. }
\label{fig:exo_fits}
\end{figure}
\begin{figure*}
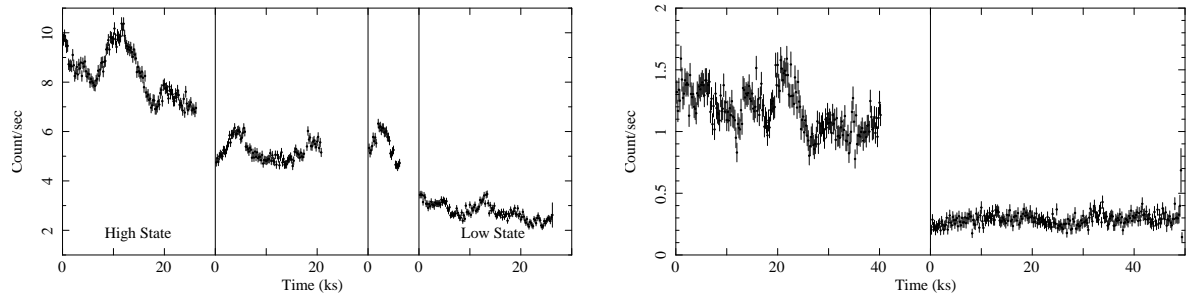

 \begin{tabular}{lr}
  \includegraphics[height=220pt,clip ,angle=270]{mrk478_lc_for_paper.ps}&
  \includegraphics[height=220pt,clip ,angle=270]{exo_lc_for_paper.ps}
 \end{tabular}
\caption{Background subtracted light curves for Mrk 478 (\emph{left}) and EXO 1346 (\emph{right}), produced with time bins of 200 seconds. The vertical lines seperate the different observations available.}
\label{fig:light_curves}
\end{figure*}

\subsection{EXO 1346.2+2645} Fitting the data with a power-law and a blackbody gave a good fit with a reduced $\chi^2$ of 1.03 for 789 d.o.f (blackbody temperature: $157\pm7$ eV, and power-law index $\Gamma=2.23\pm0.04$). Some residuals at high energies can be accounted for by adding a Gaussian line (E$ = 6.78^{+0.40}_{-0.34}$ keV, with equivalent width of $1.1^{+0.6}_{-0.4}$ keV), or a Laor line (E$=6.4^{+0.3}_{-0.4}$ keV with equivalent width of $1.4\pm0.6$ keV, $r_{\rm in}=1.3^{+25}_{-0}r_g$, and emissivity index of $2.3^{+3.2}_{-1.1}$). Also, this high energy excess can be fitted with a partial covering model (reduced $\chi^2=1.0$ for 785 d.o.f), the required column density is $6.2^{+5.1}_{-5.2}\times10^{23}$ cm$^{-2}$ with a covering fraction of $0.65^{+0.28}_{-0.18}$, the iron abundance is $2.3^{+14}_{-1.5}\times$ Solar. The high upper limit on the iron abundance indicates the data can not constrain it keV. Freezing the iron abundance at the solar value changes the column density to $11.0^{+4.0}_{-6.3}\times10^{23}$ and the covering fraction to $0.74^{+0.16}_{-0.32}$. For all fits to the high energy end, the value of $\Gamma$ does not change significantly.\\
Due to the fact that this source is faint, data from EPIC-MOS detectors were included in the analysis to improve the statistics. Fitting the whole spectrum with a reflection model gave an equally good fit with parameters consistent with the Laor line model. The power-law was also not required in this source and the data can be fitted by a reflection-only model. Similar to Mrk 478, the outer radius was fixed at 400 $r_{\rm g}$, while all other parameters were allowed to vary. The fit gave an inner radius of $r_{in}=1.6^{+0.6}_{-0.3}r_g$, and emissivity index of $3.9^{+1.1}_{-0.4}$ with an ionisation parameter of \mbox{$1202^{+120}_{-172}$ erg cm s$^{-1}$.} The best fitting model is shown in \mbox{Fig. \ref{fig:exo_fits}}.

\section{Spectral Variability}
In order to gain more insight into the nature of these two sources, light curves in different energy bands have been analysed. Fig. \ref{fig:light_curves} shows the light curves in the energy range 0.2--10 keV for the two sources. It shows that the flux of both sources change by a factor of $\sim 5$ between observations that are separated by at least a year, in addition to some small variability within single observations.
\begin{figure}
\centering
  \includegraphics[height=220pt,clip ,angle=270]{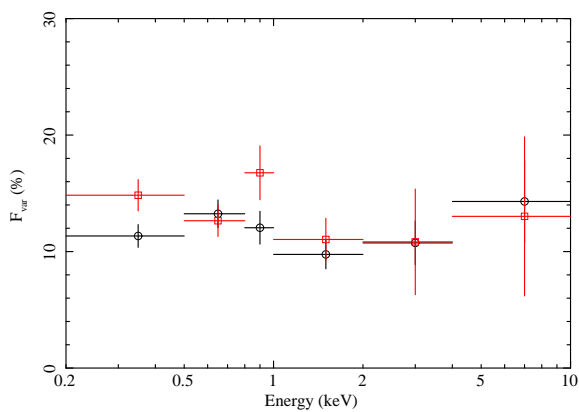}
\caption{Fractional rms variability of Mrk 478, produced with time bins of 400 seconds. \emph{Black circles:} high flux state (Obs ID: 0107660201). \emph{Red squares:} low state (Obs ID: 0005010301)}
\label{fig:mrk_rms}
\end{figure}

To study the variability in different energy bands, we used the fractional variability amplitude $F_{\rm{var}}$ (\citealt{2002ApJ...568..610E}) and hardness ratios. $F_{\rm{var}}$ was calculated following the methods described in \cite{2002ApJ...568..610E} for six energy bins. Fig. \ref{fig:mrk_rms} shows a plot of $F_{\rm{var}}$ for Mrk 478 in the ``low'' and ``high'' states, taken during the first and last observation (see light curve in Fig. \ref{fig:light_curves}).\\
It is clear that fractional variability is consistent with a constant value within each observation and between the two states. The best fitted constant lines have values of 12 and 13 percent for the high and low state, respectively. The rms spectra of AGN often have a peak at energies of about $\sim1$ keV, and these can be fully explained by a two component spectrum (e.g. \citealt{2004MNRAS.348.1415V}), with one constant reflection dominated component (RDC) and a variable power-law component (PLC). The fact that the rms spectrum is constant is in full agreement with the spectrum consisting of only one reflection component as infered from the spectral fitting.
\begin{figure}
\centering
  \includegraphics[height=220pt,clip ,angle=270]{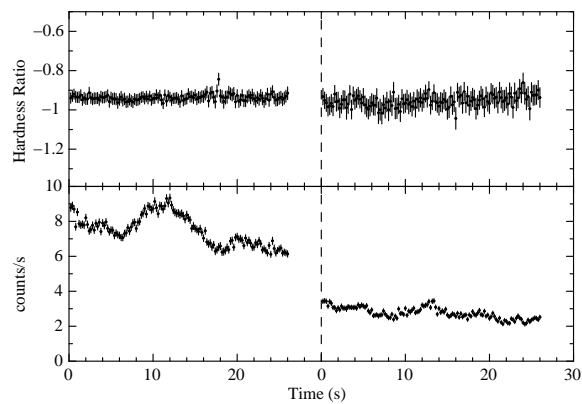}
\caption{Hardness ratio for Mrk 478 in both high and low states (Top), with the corresponding light curve (Bottom). The hardness ratio was calculated as (H-S)/(H+S), where H and S are the count rates in the 0.3--2 and 3--10 keV bands respectively.}
\label{fig:mrk_hr}
\end{figure}

To investigate this further, Fig. \ref{fig:mrk_hr} shows the hardness ratio for the low and high states (first and last observations respectively), along with the corresponding count rates. The hardness ratio was calculated using (H-S)/(H+S), with H and S being the count rates in the hard (3--10 keV) and soft (0.3--2 keV) bands respectively. It remains constant and does not respond to the flux changes. This again supports the one-component model found earlier.

\begin{figure}
\centering
  \includegraphics[height=210pt,clip ,angle=270]{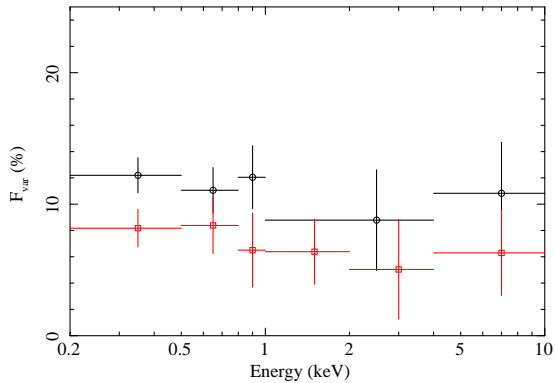}
\caption{Fractional rms variability of EXO 1345.2+2645, produced with time bins of 400 seconds. \emph{Black circles:} high flux state (Obs ID: 0097820101), and \emph{red squares} are for the low state (Obs ID: 0109070201).}
\label{fig:exo_rms}
\end{figure}

A similar trend was found in the other source (EXO 1346), the rms spectrum is shown in figure \ref{fig:exo_rms}. The fractional variability is constant in both high and low states, with the source being slightly more variable in the high state (black circles in Fig. \ref{fig:exo_rms}).

\section{Discussion}
It is clear that a simple power-law fit leaves residuals both at soft and hard energies. The soft excess can be fitted with a blackbody, with temperatures of $123^{+21}_{-30}$ and $157^{+40}_{-34}$ eV for the two sources in this study, consistent with other studies that showed that the observed range of temperatures in these sources is limited to 0.1-0.2 keV (\citealt{2004MNRAS.349L...7G}, \citealt{2006MNRAS.365.1067C}), these energies are high compared to what is expected from a standard thin accretion disc model (\citealt{1973A&A....24..337S}). A Laor line provides a good fit for the high energy residuals. However it is clear that a reflection model can fit the whole X-ray range in the these sources.

In terms of fitting parameters, in order to produce the smooth shape of the soft excess, the spectrum has to be highly blurred, implying emission from very close to the black hole. The fits presented here all point towards that, with inner radii of less than $2.5r_{\rm g}$, similar to what was found by \cite{2006MNRAS.365.1067C} for a sample of 21 sources. If the values of $r_{\rm g}$ found here are correct, then this not only implies that the black holes in these sources are rapidly spinning, but also, that strong gravitational light bending \emph{must} be occurring.

For both sources studied here, the spectra are dominated by reflection with no power-law required, which implies that the illuminating continuum is hidden from view. This is fully supported by the spectral variability analysis, that showed that the whole spectrum is varying in the same way, implying a single spectral component.\\
Although a reflection-dominated spectrum implies a deviation from symmetric geometry, it can be easily explained by a corrugated accretion flow model (\citealt{2002MNRAS.331L..35F}). The disc material is clumped into deep sheets or rings caused by disc instabilities, surrounded by a hot X-ray emitting corona. The hard X-ray continuum source is hidden for most observers and only the reflection off the sheet is observed. Alternatively, it can be explained in terms of strong light bending effects in the vicinity of the black hole (\citealt{2003MNRAS.344L..22M}, \citealt{2004MNRAS.349.1435M}). In this model, emission from the illuminating source is bent back towards the black hole, while the reflected photons can easily escape towards the observer.

There has been lately a controversy over the exlanation of the spectral features of some NLS1. In particular the sharp drop at $\sim7$ keV in 1H0707-495 (\citealt{2002MNRAS.329L...1B}), which was explained either by reflection or partial covering. It is clear that this feature is not seen in the sources presented here, and both models give an equally good fit to the data, but based on the spectral variability, the reflection model is favoured. The sources show flux variability of up to a factor of 5 between epochs, with no change in the spectrum. In a source covered by a line-of-sight clouds, this requires that the observed variability is due to variations in the intrinsic source not the clouds covering it (because the parameters describing the coverer did not change). By causality argument, the intrinsic source is less than a lighthour across, and the partial coverer needs to be even smaller, yet has to give the same covering fraction years later, this seems completely implausible to us. The reflection model on the other hand, fits the whole spectrum vey well and is physically motivated, with the changes between epochs explained by changes in the ionisation of the reflector.

\section*{Acknowledgements}
AZ thanks the Cambridge Overseas Trust and STFC. ACF thanks the Royal Society for support.

\bibliography{main}

\label{lastpage}

\end{document}